\documentclass[a4paper,fleqn,usenatbib]{mnras}

\usepackage[T1]{fontenc}
\usepackage{ae,aecompl}

\usepackage[toc,page]{appendix}

\usepackage{graphicx}	
\usepackage{amsmath}	
\usepackage{amssymb}	

\title[Constraining disks]{Constraining protoplanetary disks with exoplanetary dynamics: Kepler-419 as an example}

\author[Ali-Dib \& Petrovich]{
Mohamad Ali-Dib$^{1,3}$\thanks{E-mail: malidib@astro.umontreal.ca} \& Cristobal Petrovich$^{2,4}$ \\
$^{1}$Institut de recherche sur les exoplan\`etes, Universit\'e de Montr\'eal,
2900 boul. \'Edouard-Montpetit, Montr\'eal, H3T 1J4, Canada\\
$^{2}$Steward Observatory, University of Arizona, 933 N. Cherry Ave., Tucson, AZ 85721, USA\\
$^{3}$Centre for Planetary Sciences, University of Toronto Scarborough, Toronto, Ontario M1C 1A4, Canada\\
$^{4}$Canadian Institute for Theoretical Astrophysics, 60 St. George St., Toronto, ON M5S 3H8, Canada
}

\date{Accepted XXX. Received July 2020.}

\pubyear{2020}

\begin{document}
\label{firstpage}
\pagerange{\pageref{firstpage}--\pageref{lastpage}}
\maketitle

\begin{abstract}
We investigate the origins of Kepler-419, a peculiar system hosting two nearly coplanar and highly eccentric gas giants with apsidal orientations librating around anti-alignment, and use this system to place constraints on the properties of their birth protoplanetary disk. We follow the proposal by \cite{petrovich} that these planets have been placed on these orbits  as a natural result of the precessional effects of a dissipating massive disk and extend it by using direct N-body simulations and models for the evolution of the gas disks, including photo-evaporation. 
Based on a parameter space exploration, we find that in order to reproduce the system the initial disk mass had to be at least 95 M$_{\rm Jup}$ and dissipate on a timescale of at least 10$^4$ yr. This mass is consistent with the upper end of the observed disk masses distribution, and the dissipation timescale is consistent with photoevaporation models. We study the properties of such disks using simplified 1D thin disk models and show that they are gravitationally stable, indicating that the two planets must have formed via core accretion and thus prone to disk migration. We hence finally investigate the sensitivity of this mechanism to the outer planet's semi major axis, and find that the nearby 7:1, 8:1, and 9:1 {mean-motion resonances}  can completely quench this mechanism, while even higher order resonances can also significantly affect the system. Assuming the two planets avoid these high order resonances and/or close encounters, the dynamics seems to be rather insensitive to planet $c$ semi major axis, and thus orbital migration driven by the disk.

\end{abstract}

\begin{keywords}
planets and satellites: formation -- planets and satellites: gaseous planets -- planet-disc interactions
\end{keywords}


\section{Introduction}
Kepler-419 is a two gas giants system with well characterized architecture. The two planets $b$ and $c$ are respectively 2.77 and 7.65 Jupiter masses, orbiting at 0.374 and 1.697 AU, with eccentricities of 0.81 and 0.18, in apsidally anti-aligned orbits ($\varpi{_b}$ - $\varpi{_c}$ $\sim$ 180 $\deg$) \citep{ford, dawson, almenara}. The origins of these unique orbits merit an explanation. \cite{petrovich} (PWA19) showed using secular theory (approximate orbit-averaged equations of motion) that the system could have originated in the inner gap of a slowly dissipating massive disk that forced the apses to anti-align through its precessional effects. {The general dynamics and eccentricity evolution of exoplanets due to disk dispersal were initially explored by \cite{Nagasawa}.} In this mechanism, the system needs to start with an angular momentum deficit (AMD) for planet $c$, that is transferred in the process to planet $b$. { PWA19 proposed that this initial AMD might be due to either planet–disk interactions where the outer Lindblad resonances can increase the planet's eccentricity \citep{bitsch1}, or from planet–planet scattering \citep{lega}. It is unlikely however that scattering alone can lead to an anti-aligned and nearly coplanar system \citep{barnes,Chatterjee}, hence the need for the disk dispersal mechanism. We refer the reader to PWA19 for further background discussions. }

An alternative explanation was proposed by \cite{jackson}, who showed using N-body integrations that for a small region of parameter space, the presence of an undetected third planet could excite planet $b$'s eccentricity periodically without distabilizing the system. In this paper we build on and extend the work of PWA19. We first verify the accuracy of their results using N-body integrations and {do a parameter study over disk mass and dispersal timescale to constrain the values allowing the formation of Kepler-419 (sections \ref{pef} and \ref{popsyn}).} We then study the properties of these disks and compare them against observations to verify the realism of this idea (section \ref{disk}). We finally study the effects of changing planet $c$'s semi-major axis to understand the sensitivity of the system (and formation mechanism) to this parameter (section \ref{csm}).

\section{Numerical setup}
All simulations in the work were done using the \texttt{REBOUND} N-body integrator \citep{r4,r2,r3}, along with its \texttt{REBOUNDx} add-ons package \citep{r1}. 
Simulations were mostly done using the symplectic Wisdom-Holman integrator \texttt{WHFAST}, unless otherwise is explicitly stated. This is because we are only interested in the evolution of stable systems that do not undergo close encounters. The system is integrated with a timestep equal to 0.025 $\times$ the smallest orbital period in the system. All simulations are run for $6\times10^6$ yr.  The disk potential is implemented following \cite{tremaine} eq. 2.156, leading to the radial acceleration:
\begin{equation}
\frac{F_{D}(r)}{m}=\frac{4 G}{r} \int_{0}^{r} d a \frac{a}{\sqrt{r^{2}-a^{2}}} \frac{d}{d a} \int_{a}^{\infty} d r^{\prime} \frac{r^{\prime} \Sigma\left(r^{\prime}\right)}{\sqrt{r^{\prime 2}-a^{2}}}
\end{equation}
where we are setting to 0 the $z$ component of the potential, and hence are not considering its effects on the inclinations of the system. In fact we treat the system purely as 2 dimensional, in contrast with the 3D treatment of PWA19. 

The disk force is implemented using the \texttt{add$\_$custom$\_$force} method of \texttt{REBOUNDx}. 

For all cases, we initiate the system with a 1.39 $M_\odot$ star, and two planets $b$ and $c$ of 2.77 and 7.65 $M_J$. Their respective initial eccentricities are set to 0.05 and 0.4 (following PWA19), while inclinations and Longitudes of ascending nodes are set to 0. We again follow PWA19 by initially setting $\omega_{\mathrm{b}}-\omega_{\mathrm{c}}=60^{\circ}$ 

\section{Kepler-419: planets fully embedded in a photoevaporating disk}
\label{pef}

For the system's secular dynamics to evolve as suggested by \cite{petrovich}, a main requirement is for the two planets to be located inside a common gap, with a massive disk beyond their orbits. One possible physical mechanism for this scenario is a photoevaporating disk, where the planets are initially fully embedded in a disk with no cavity, then photoevaporation slowly carves up a gap in the inner disk. 

Here we test this hypothesis numerically using our setup, and assuming a disk surface density profile that evolves as a function of time as:

\begin{equation}
\label{sfe}
\Sigma(r,t) = \Sigma_0(r) - \dot{\Sigma}_{w}\times \Delta t
\end{equation}
where $\Sigma_0(r)$ is defined through the following functional form:
\begin{equation}
\label{sigma}
\Sigma(R)=\Sigma_{0}\left(\frac{r}{r_{\text{in}}}\right)^{-\gamma}
\end{equation}
where $\Sigma_{0}$ is one of the main parameters we vary, $r_{\text {in}}$ is set to 0.05 AU, and $\gamma$ is fixed at 1.5. The outer edge of the disk is always kept at 50 AU. The photoevaporation rate $\dot{\Sigma}_{w}$ is defined following the functional fit to hydrodynamic simulations of \cite{owen} :

\begin{equation}
\begin{aligned} \dot{\Sigma}_{w}(y)=&\left[\frac{a_{2} b_{2} \exp \left(b_{2} y\right)}{R}+\frac{c_{2} d_{2} \exp \left(d_{2} y\right)}{R}+\frac{e_{2} f_{2} \exp \left(f_{2} y\right)}{R}\right] \\ & \times \exp \left[-\left(\frac{y}{57}\right)^{10}\right] \end{aligned}
\end{equation}
and $\dot{\Sigma}_{w}(y<0)=0$, where
\begin{equation}
{y=0.95\frac{\left(R-R_{\text {hole }}\right)}{1\mbox{AU}}\left(\frac{M_{*}}{1 \mathrm{M}_{\odot}}\right)^{-1}} 
\end{equation} 
and the dimensionless constants are $a_{2}$=-0.438226, $b_{2}$=-0.10658387, $c_{2}$=0.5699464 $d_{2}$=0.010732277, $e_{2}$=-0.131809597, $f_{2}$=-1.32285709.
 
We use R$_{hole}$=0.05 AU. We normalize this photoevaporation rate by $10^{8}$, to get a total mass loss of $6\times 10^{-8} M_\odot/yr$. 

\begin{figure*}
\begin{centering}
	\includegraphics[scale=0.25]{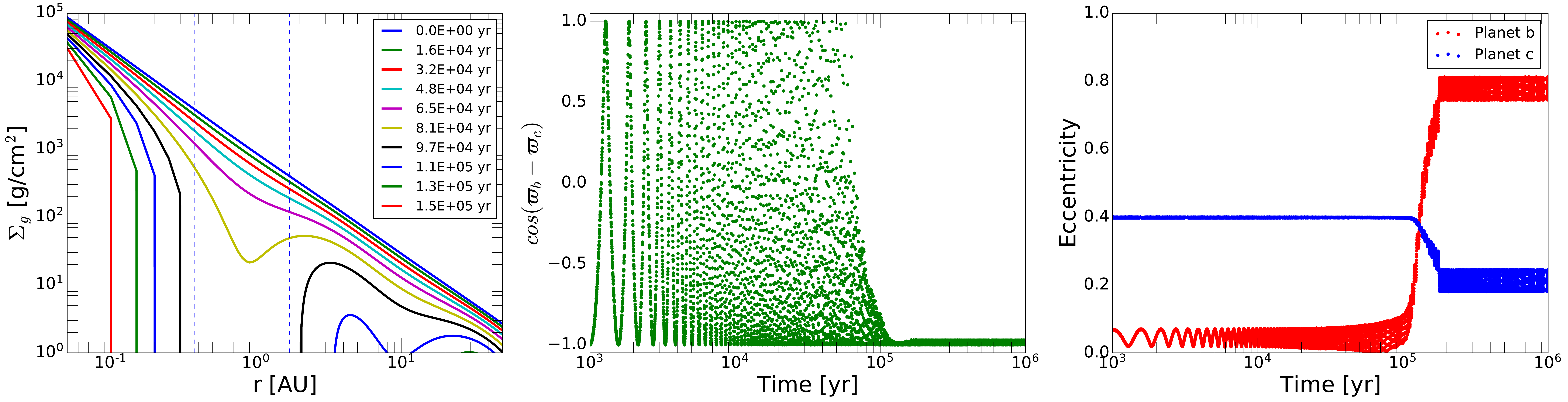}
   \caption{The evolution of a two planets system into Kepler-419 in a photoevaporating disk. Left: Time evolution of the gas surface density profile of a disk governed by eq. \ref{sfe}. Center: The apses of the two planets are forced into anti-alignment by the disk's precession, and then remain in this state due to its adiabatic dispersal. Right: Starting with an AMD, the eccentricity of planet $b$ increases from a near zero to $\sim$ 0.8, while that of planet $c$ decreases from the initial 0.4 to $\sim$ 0.2.  }
    \label{fig:peall}
    \end{centering}
\end{figure*}

 The system's evolution is shown in Fig. \ref{fig:peall}. The disk starts with a mass of $\sim 10$ M$_J$. Photoevaporation then disperses the disk on a timescale of $\sim 10^5$ yr, opening a gap spanning the region between the two planets after $\sim 8\times 10^4$ yr. At this point $\sim 5$ M$_J$ of gas is remaining outside the orbit of planet $c$. Before the gap opening, planet $c$ eccentricity remains constant while planet $b$'s oscillate with moderate amplitude, consistent with the Laplace-Lagrange solution for two planets orbiting a star, with the axi-symmetric disk having no effect as the two planets are still fully embedded. When the gap is opened at $\sim$ t=$8\times 10^4$ yr however, angular momentum exchange proceeds with planet $b$'s eccentricity increasing quickly to 0.8, and planet $c$'s decreasing to 0.2, values consistent with current observations of Kepler-419. Finally, the gap opening is followed by the anti-alignment of the planets apses, the second major dynamical characteristic of Kepler-419.

{In reality however, this approach is not truly self-consistent since, as shown in section \ref{disk}, planets this massive will open deep gaps in the disk that are wide enough to merge, quickly clearing up the disk regions interior to the outer planet. In section \ref{popsyn} we consider a more realistic, although simpler, setup where the planets start embedded in a common gap. The dynamics considered in the current section however might still be relevant for sub-Neptunes not capable of fully carving gaps. For more sophisticated models of disk evolution with planet-carved gaps we refer the reader to the recent work by \cite{toliou}. }

\section{Kepler-419: parameters study}
\label{popsyn}
In this section we investigate the disk mass and dispersal timescales needed for a 2 planets system to evolve into Kepler-419 like configuration.
We hence generate and evolve a large number of systems while varying two parameters: 1- the disk mass through $\Sigma_0$ in eq. \ref{sigma}, and 2- the disk's dispersal timescale $\tau_{\mathrm{d}}$. Therefore, instead of using eq. \ref{sfe} as above, we simply the scheme by assuming the disk mass to decrease as $M_{disk}^{t0} \exp\left({-t/\tau_d}\right)$. This implies that the disk is dissipating simultaneously at all radii, rather than inside-out as suggested by photoevaporation models. 

{This is however justifiable, since, within the simplifications of the model (axisymmetric disk, no feedback from the planet onto the disk), the dynamics we are interested in has a stronger dependence on the disk's total mass, than on its distribution. This is shown in PWA19's eqs. 9 and 11 where  for a minimal mass solar nebula density profile: $\dot{\varpi}_p \propto M_{disk}/\sqrt{r_{out}}$ for a given $r_{in}$.}


We set $r_{in}$ in eq. \ref{sigma} to 1.5$\times a_c$ AU with $\Sigma(r<r_{in}) = 0$ to emulate a gap, and the semi-major axis of planets $b$ and $c$ to respectively 0.374 and 1.697. 

We try disk masses of 1, 20, 40, 50, 75, 100, and 200 $M_{\rm Jup}$. The higher values are probably unrealistic as such massive disks can be prone to gravitational instabilities, but we include them as limiting cases. We explore disk dispersal timescales $\tau_d$ of $10^4$, $10^5$, and $10^6$ yr.

\subsection{Disk mass and dispersal timescale}
\begin{figure*}
\begin{centering}
	\includegraphics[scale=0.25]{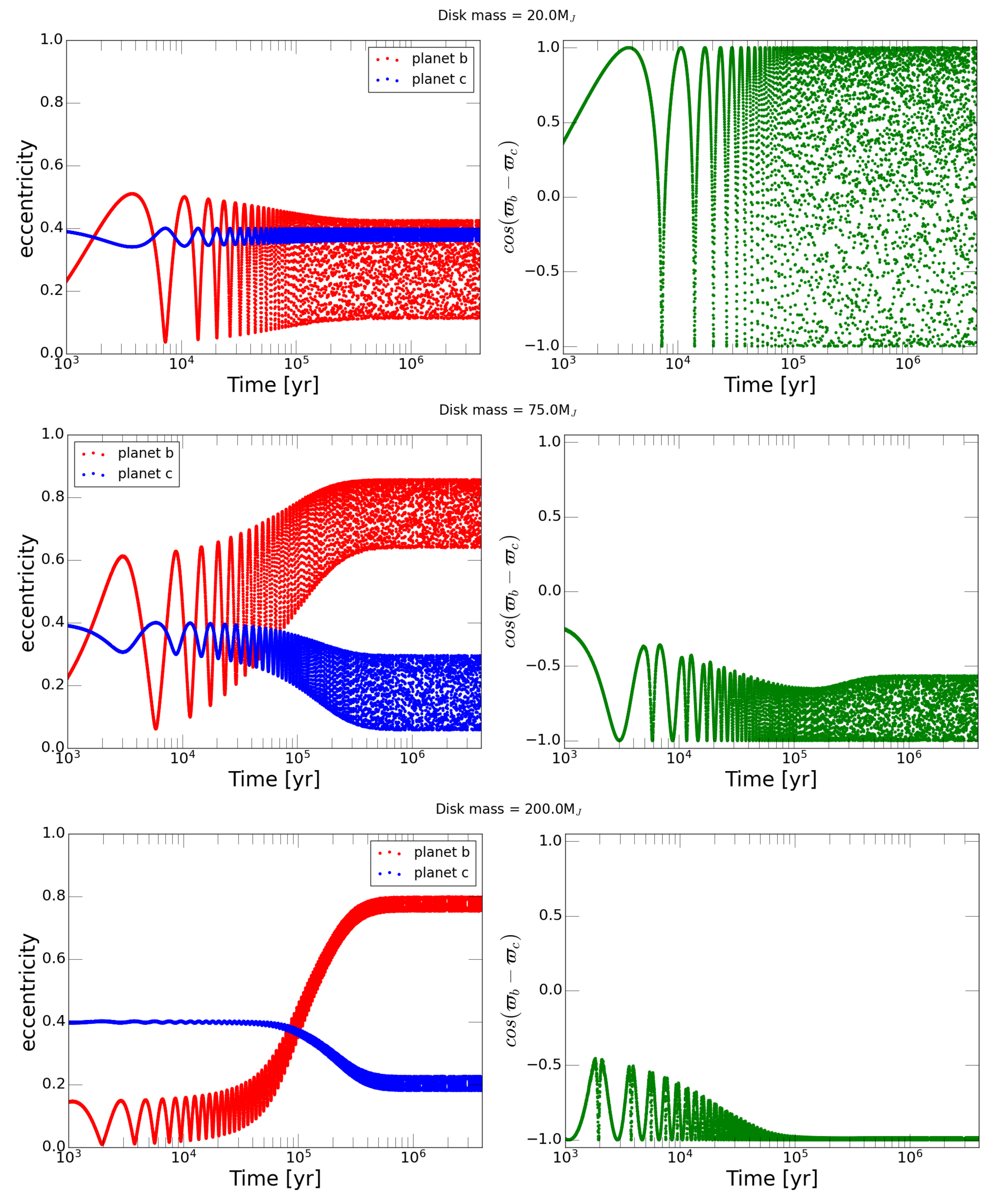}
   \caption{Examples for typical behaviors observed in our Kepler-419 like setups, for three disk mass values, and $\tau_d = 10^5$ yr. Top: the disk mass is too low to have any effects on the system evolution. Middle: Transition case where the disk mass forces the planets apses into anti alignment but with noticeable oscillation amplitudes. Bottom: Massive disk leading to Kepler-419 like system with low oscillation amplitudes.}
    \label{fig:main1b}
    \end{centering}
\end{figure*}

In Fig. \ref{fig:main1b} we show the time evolution of the planets' eccentricities and apses for three representative cases, with the full results for all cases shown in appendix Figures \ref{fig:app1} and \ref{fig:app2}.

In this plot we identify multiples distinct regimes. For $M_d = 1 \ \text{and} \ 20 \ M_{\rm Jup}$, the disk mass is small enough for the disk-planet potential to be weak compared to that planet-planet interaction potential. Hence this case is equivalent to a three-body problem with a central star and two significantly less massive planets where the eccentricities are secularly forced and the the apses circulate. This problem is well described by the classical Laplace-Lagrange secular theory \citep{murray}. 

On the other extreme end, for the massive disks with $M_d \geq 75 \ M_{\rm Jup}$, we recover the results of PWA19 remarkably. In all of these cases the apsidal precession of the planets is dominated by the disk, rather than the planet-planet secular interactions. The apses therefore always evolve into liberating around anti-alignment, while the AMD is transferred from planet $c$ to planet $b$, giving eccentricities consistent with Kepler-419 as seen again in Fig. \ref{fig:main1b}. Finally, for the intermediate cases of $M_d \sim 40-50 \ M_{\rm Jup}$, we do see possible evolution towards (sometimes transient) anti-aligned apses, however the amplitude of the oscillations are very large and thus we consider this case incompatible with Kepler-419. The same can be said for the eccentricities. 

We identify the minimal disk mass necessary for a Kepler-419 like system to be around 75 M$_{\rm J}$, a value few times higher than that used in section \ref{pef} where the system did evolve into K419. The two setups however are not fully equivalent. { In section \ref{pef} the planets start fully embedded in the disk, and hence the precession rate of the planets is retrograde (e.g., \citealt{rafikov}) and dominated by the inner planet ($|\dot{\varpi}_{\rm in}|\propto M_{\rm disk} n_{\rm in}$). In both cases, we have $\Delta  \dot{\varpi}\equiv \dot{\varpi}_{\rm out}-\dot{\varpi}_{\rm in}>0$ initially, allowing the resonance to be crossed ($\Delta\dot{\varpi}=0$) in a similar fashion. However, the mass of the disk required to cross the resonance is lower in the embedded case roughly by a factor of $n_{\rm in}/n_{\rm out}=P_{\rm out}/P_{\rm in}\sim 9$, which is roughly consistent with the ratio of the minimal disk masses ($\sim 75 M_J/ 10 M_J = 7.5 $) between the fully embedded simulations and this case. }


It is interesting that Kepler-419 is recovered even for disks dispersing on a $10^4$ yr timescale, an order of magnitude lower than the value assumed by PWA19.  This is reassuring since in photoevaporation models, once an inner cavity has been carved (which is our starting assumption), disks usually proceed to disperse inside-out very quickly due to direct stellar irradiation. 
This puts a constraint on the amount of ``adiabacity'' necessary for this mechanism to operate. Notice that, as one would expect, for a fixed disk mass, longer disk dispersal timescales lead to the same end results, but over longer time. On the other hand for fixed $\tau_d$ but increasing the disk mass, the amplitude of the secular oscillations around the equilibrium values of the eccentricity and apses decreases.

\subsection{Properties of the natal disk}
\label{disk}
In this section we analyze the properties of our lowest mass disk compatible with Kepler-419, that is 75 $M_J$ ($\sim$0.075 M$_\odot$). First we compare this disk to observations. In a recent ALMA survey of the Lupus complex, \cite{ansdell1} found dust masses ranging from $\sim 0.3$ to 100 M$_\oplus$, while their gas masses where mostly below Jupiter mass, and almost all of them below the MMSN (10$^{-2}$ M$_\odot$, 10 M$_J$). \cite{pascucci} and \cite{long} also used ALMA to measure the dust mass in the Chamaeleon I star-forming region, and found a mass range consistent with \cite{ansdell1}. Gas masses as measured in the infrared (with HD lines) seem to be higher, where \cite{mclure} for example constrained the masses of GM Aur and DM Tau to respectively 2.5-20.4$\times 10^{-2}$ and 1.0-4.7$\times10^{-2}$ M$_\odot$. \cite{bergin} on the other hand found a lower mass limit of TW Hya around 0.05 M$_\odot$ (50 M$_J$). Our lowest mass disk hence lies towards the upper end of these distributions. Note that estimating disk gas mass from CO lines is  problematic, as it depends on the assumed CO/H$_2$ ratio, that is affected by the complex physical-chemistry of CO. \cite{yu} for example found that CO observations underestimate gas mass by an order of magnitude. The values reported above are hence probably lower limits in most cases. Note that since dust mass is measured using the sub-millimeter continuum flux that is insensitive to solids larger than $\sim$ cm, dust masses are also lower limits that does not account for the total solids budget. Dust mass is hence a bad tracer for the total disk's mass, even when the stellar metallicity is known.  \\

To fully understand the Kepler-419 birth protoplanetary disk, we need to account for the planets mass as well. This increases our minimal mass disk to 95.07 $M_J$ (0.094 $M_\odot$), still around the uppermost limits of the observed population. Assuming ISM dust/gas ratio of 0.01, the minimal mass dust disk is $\sim$ 300 M$_\oplus$, a factor of few times above the upper limits of the measured values. On the M$_{*}$-M$_{dust}$ diagrams of \cite{ansdell2}, our minimal mass disk fit on the curves found for the younger disks in Taurus, Lupus, and Cham I. 

{ Our minimal mass K419 disk can moreover be compared to transitional disks with a large inner cavity. These disks are an intermediate step between photoevaporating protoplanetary disks and debris disks. A \textit{Spitzer} survey of $\sim$ 150 transition disks by \cite{vander} found that $\sim 7$\% of there sample has a total mass $\geq$ 100 M$_J$, even though there is a much higher fraction with masses between 10-100 M$_J$. These masses were obtained through (dust) SED fitting by \texttt{RADMC-3D} radiative transfer models, assuming ISM dust to gas ratio and account for dust growth using the prescription of \cite{andrews}. A \textit{Herschel} data analysis of Chamaeleon I by \cite{ribas} found that SZ Cha and CS Cha have respectively M$_{dust}$ = 10$^{-3.4}$ and 10$^{-3.8}$ $M_\odot$, implying roughly  10$^{-1.6}$ $M_\odot$ total mass. This is a factor of 3 times less than our minimal disk. Finally, a recent complete ALMA survey of the Lupus Star-forming Region by \cite{vander2} found that 3 out of 11 disks have dust masses consistent with our minimal mass disk. Therefore, transition disks massive enough to form K419 are consistent with observations. Note that, strictly speaking, transition disks masses should be compared only to the minimum disk mass excluding the actual planets (75 $M_J$), while protoplanetary disks masses should be compared to the minimal disk in addition to the planets (95 $M_J$). The significant uncertainties on the disks masses, due to both observational error bars and the modeling assumptions, render this distinguishment unnecessary however. }

Now we focus on the possible formation pathways for the Kepler-419 planets. Since a very massive disk is needed to form the system, we investigate whether such a disk is gravitationally stable. We hence construct a standard radiative steady-state thin disk \citep{pringle}, as described in \cite{alidib}. This gives:

\begin{eqnarray}
    T_{d,\mathrm{rad}} &=& T_s^{4/5} \left({\kappa\mu\Omega \dot M\over 6\pi \alpha k_B \gamma}\right)^{1/5}\nonumber\\
    &=& 373\ {\rm K}\  r_{\rm AU}^{-9/10} \alpha_{-2}^{-1/5} \dot M_{-7.5}^{2/5}\nonumber\\ &&\times\left({M_\star\over M_\odot}\right)^{3/10}\left({\kappa\over {\rm cm^2\ g^{-1}}}\right)^{1/5}
\end{eqnarray}
where $\alpha_{-2}=\alpha/0.01$ and $\dot M_{-7.5} = \dot M/10^{-7.5}\ M_\odot\ {\rm yr^{-1}}$. The density is then 
\begin{eqnarray}
    \rho_{d, \mathrm{rad}} &=& 1.7\times 10^{-10}\ {\rm g\ cm^{-3}}\ r_{\rm AU}^{-33/20}\alpha_{-2}^{-7/10}\dot M_{-7.5}^{2/5} \nonumber\\
    &&\times \left({M_\star\over M_\odot}\right)^{11/20}\left({\kappa\over {\rm cm^2\ g^{-1}}}\right)^{-3/10}.
\end{eqnarray}

\begin{figure*}
\begin{centering}
	\includegraphics[scale=0.15]{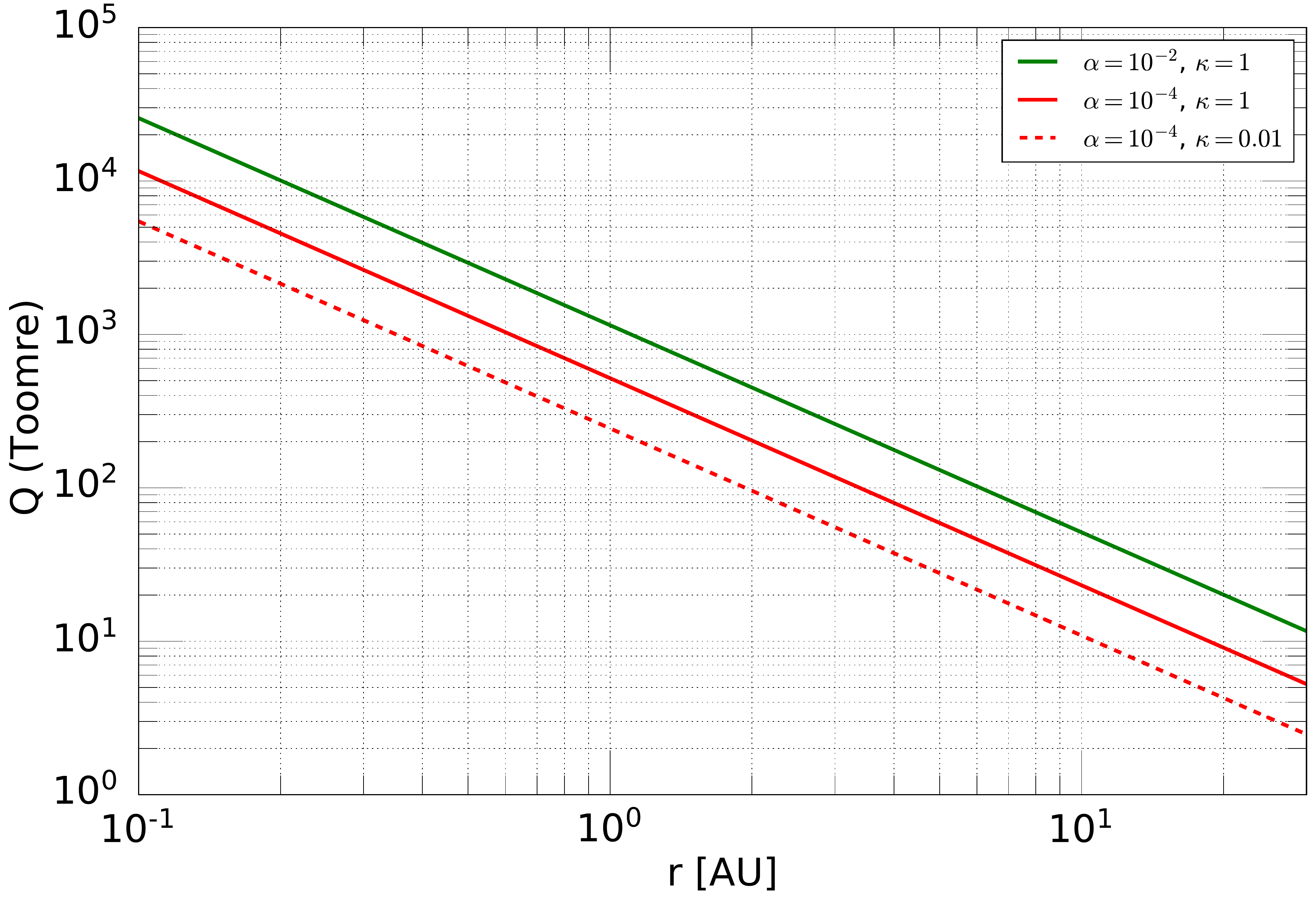}
	\includegraphics[scale=0.15]{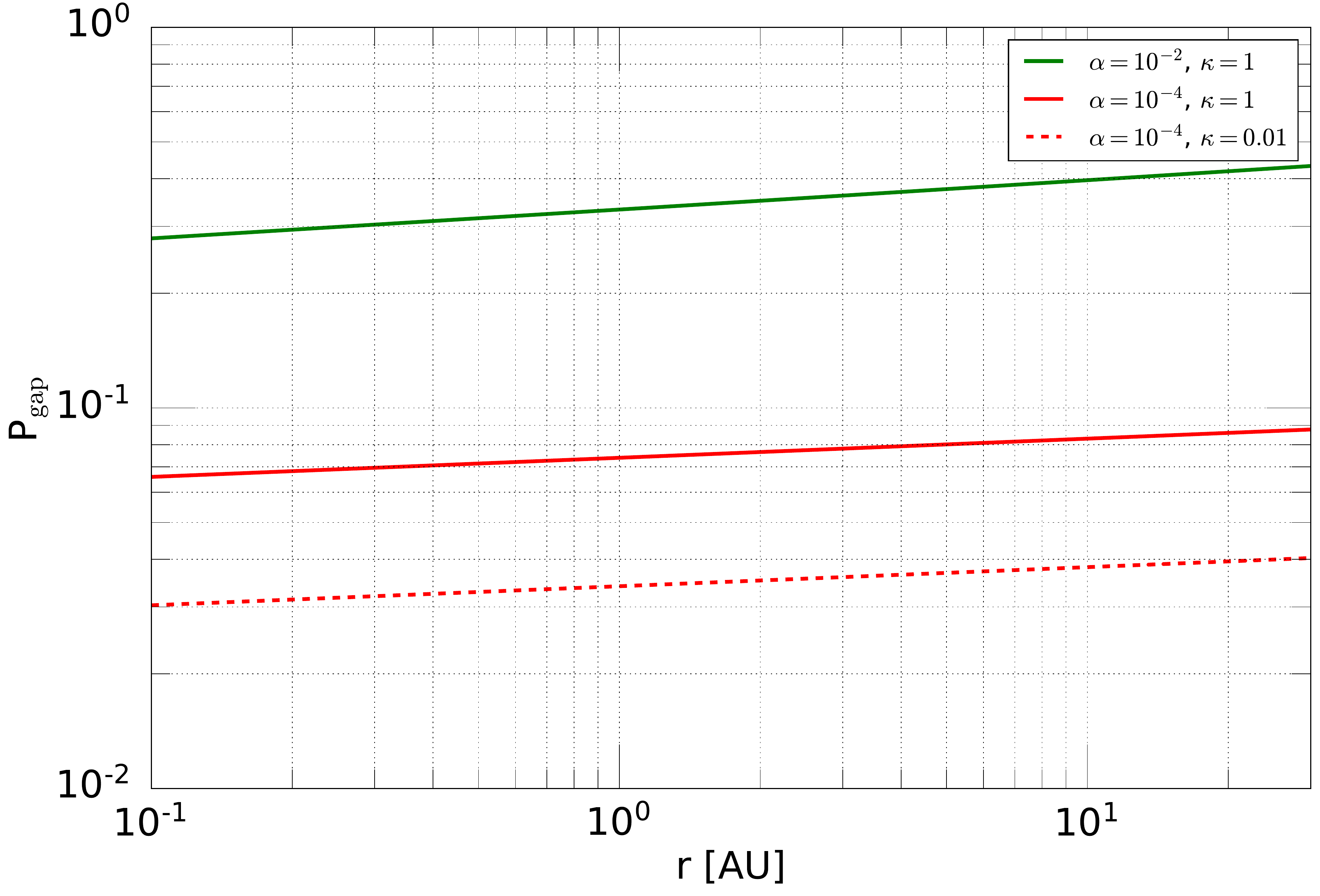}
	\includegraphics[scale=0.15]{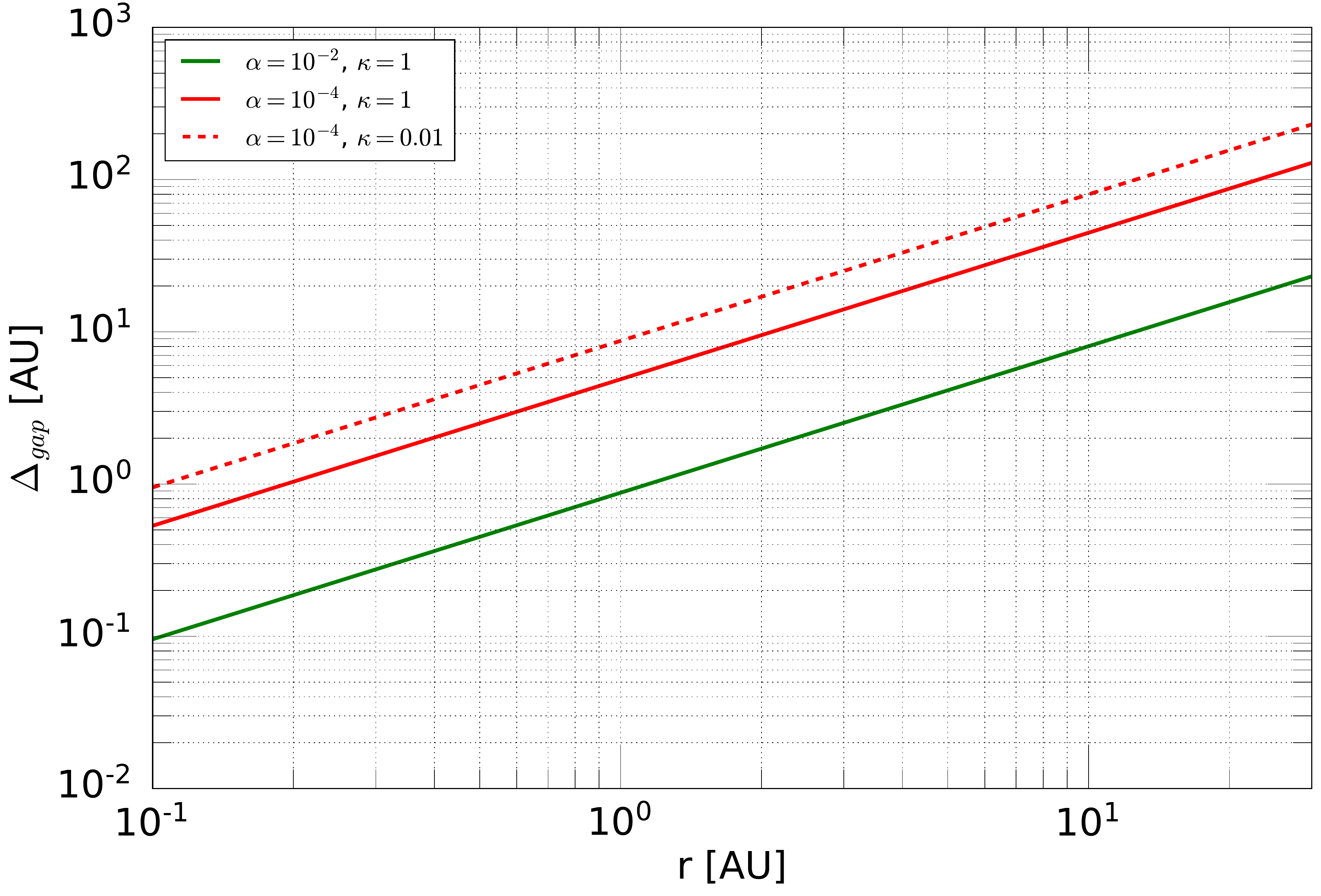}
   \caption{Left: The Toomre stability parameter $Q$ for our minimal mass Kepler-419 disk, for different values of turbulent viscosities $\alpha$ and disk opacity $\kappa$. In all cases the disk is fully stable against gravitational collapse in the entire planet formation region. Center: The value of the gap opening criteria function $P_{gap}$ as defined in eq. \ref{eqpgap}, for a 2.77 M$_J$ planet. $P_{gap}$ is significantly lower than 1 throughout the disk, indicating that this planet is capable of carving a deep cavity anywhere. Right: The width of the gap opened by a 2.77 M$_J$, as calculated through eq. \ref{eqwidth}. For all parameters, a gap carved by a planet at 1.7 AU is deep enough to reach the inner Kepler-419 planet no matter its mass, allowing the two planets to be embedded inside a common gap, without the need for photoevaporation. }
    \label{fig:maingap}
    \end{centering}
\end{figure*}

We hence control three disk parameters: turbulence viscosity parameter $\alpha$, the disk opacity $\kappa$, and its accretion rate onto the star $\dot M$. We try $\alpha=10^{-2}$ and $10^{-4}$, in addition to $\kappa=1$ and 0.01. For each case we change the $\dot M$ value to get the total disk mass we need. 
We finally calculate the Toomre parameter:
\begin{equation}
Q \equiv \frac{c_{s} \Omega}{\pi G \Sigma}
\end{equation}
where $Q < 1$ for gravitationally unstable disks. We plot $Q$ in Fig. \ref{fig:maingap} (left hand panel) for our three disk models: highly turbulent with high opacity (``hot'' disk), and weakly turbulent with both high and low opacity. 

We find that, in all cases, $Q \gg 1$ everywhere in the planet forming regions of these disks inside 30 AU. These massive disks are hence gravitationally stable under our simplified assumptions. While it is conceivable that gravitational collapse can take place in the outermost parts of the disk (where it is usually thought to operate), a gas giant will only undergo the slow type II migration on a timescale of $r^2/\nu$. For $r=30 AU$ and $\nu \sim 10^{15} cm^2/s$, $t_{mig} \sim 6\times10^6$ yr, which is comparable to the disk lifetime. 
Assuming these calculations stands for more sophisticated disk and migration models with proper radiative transfer, this imply that the Kepler-419 planets probably formed via core accretion.  \\

It is also of interest to check whether the Kepler-419 planets are capable of carving gaps in such massive disks. This is an important self consistency check since an inner gap in the disk embedding both planets is a prerequisite for the secular dynamics we are considering, and this can be an alternative to the photoevaporation carved gap discussed in section \ref{pef}. \cite{crida} showed that for a planet embedded in a disk to open a gap, the following condition needs to be satisfied:
\begin{equation}
\label{eqpgap}
P_{gap} \equiv  \frac{3}{4} \frac{H}{R_{H}}+\frac{50}{q \mathcal{R}} \lesssim 1
\end{equation}
where $R_H$ is the planet's Hill radius, $q=M_p/M_s$, and $\mathcal{R}=r_{p}^{2} \Omega_{p} / \nu$ is the Reynolds number.
In Fig. \ref{fig:maingap} (center) we plot this quantity for our disk models, and $M_p=2.77$ M$_J$. We find the condition to be satisfied throughout the disk for both $\alpha=10^{-4}$ and $10^{-2}$. This imply that the Kepler-419 planets can indeed open gaps in these massive disks.

We finally calculate the gaps' widths following \cite{width}:
\begin{equation}
\label{eqwidth}
\frac{\Delta_{\text {gap }}}{R_{p}}=0.41\left(\frac{M_{p}}{M_{*}}\right)^{1 / 2}\left(\frac{h_{p}}{R_{p}}\right)^{-3 / 4} \alpha^{-1 / 4}
\end{equation}
with the results shown in Fig \ref{fig:maingap} (right). Even for $\alpha=10^{-2}$, the gaps are found to be wide enough to merge, allowing the two planets to coexist inside one common gap. This imply that a photoevaporation driven inner disk gap is not the only possible formation channel for Kepler-419, as the planets are massive enough to exist within a common gap anyway. 


\subsection{Sensitivity of the results to $P_c$}
\label{csm}

\begin{figure*}
\begin{centering}
	\includegraphics[scale=0.5]{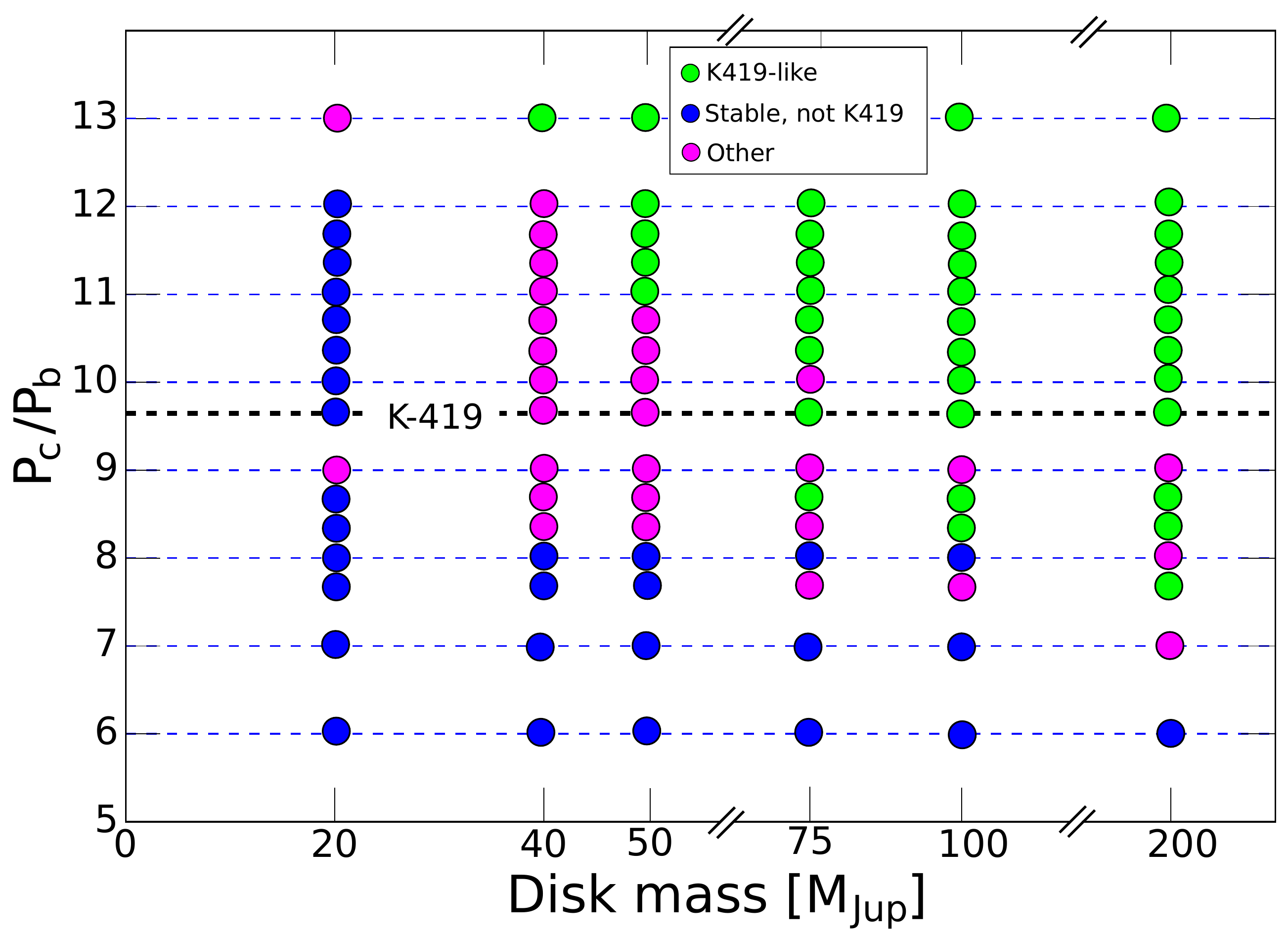}
   \caption{The effects of relatively small changes in the period (thus semi major axis) of planet $c$. { Blue} circles are systems that become stable, with little to no eccentricity or apsidal evolutions, and low amplitude secular oscillations. { Green} circles represent cases that evolve into Kepler-419 like systems where the apses liberate around anti-alignment.
   Finally, {Pink} circles on the other hand are systems that does not fit either of the previous two categories. Notice the multiple breaks in the x-axis scale. }
    \label{fig:main2a}
    \end{centering}
\end{figure*}

If the Kepler-419 planets formed via core accretion as argued above, then it is likely they underwent disk migration at some point in their history. This is especially true for massive disks, since the timescale of type I migration is $\propto \Sigma_g$. In this section we hence explore the dependence of Kepler-419's architecture on the semi-major axis / period of planet $c$, that we now free as a parameter. In Fig. \ref{fig:main2a} we show the end-results of simulations where we changed the period of planet $c$ for values ranging between 6 and 13 $\times$ P$_b$. The actual value for Kepler-419 is $P_c$ = 9.6 $P_b$. We use the same disk mass range (20 to 200  M$_{\rm J}$), and a fiducial dispersal timescale of 10$^5$ yr.

Multiple distinct dynamical regimes can be identified in this plot.

Green circles in Fig. \ref{fig:main2a} represent systems that evolved into a Kepler-419 like architecture, with anti-aligned apses, and the observed eccentricities of both planets. These are mainly cases with $P_c$ > 8$\times P_b$ and $M_d \geq$ 50 M$_{\rm J}$. We moreover notice that the minimum disk mass needed for the system to follow this evolution channel decreases with $P_c$:  while 75 M$_{\rm J}$ are needed for the $P_c$ = 9.6 $P_b$, only 40 M$_{\rm J}$ are needed for $P_c$ = 13 $P_b$. This can be readily understood within the framework of PWA19's analytic model where planet $c$' precession timescale due to the disk is $\propto P_c/M_d$.

Blue circles on the other hand are systems that are either completely stable (with no apsidal liberation or AMD exchange between the two planets), or with minor dynamical evolution that does not evolve the system into Kepler-419 (mostly moderate amplitude secular oscillations). This is the case for low disk masses that are unable to significantly affect the system's dynamics, reducing the setup into a classic 3-body problem with an inner test particle and external perturber.
More interestingly, this is also the case for some of the MMR cases, even for high disk masses. Prominent examples are the 6:1, 7:1, and 8:1 resonance for almost all disk masses.

Pink circles are systems that undergo significant dynamical evolution without transforming into K419. This is an umbrella for a plethora of different behaviors. For $M_d$ = 40 M$_{\rm J}$ for example, while planet b's eccentricity does increase significantly while that of planet c is decreasing, the amplitudes of the eccentricities secular oscillations is very large. Moreover, while the planets' apses do anti-align, this state in these cases is transient. An intriguing case is that of the 9:1 MMR where no AMD exchange takes place between the planets at all, but their apses do liberate stably around anti-alignment. For periods slightly smaller or larger than this exact commensuration however, the system evolves cleanly into Kepler-419. The 10:1 MMR is also a unique case, where the planets' follow a dynamically very ``noisy'' Kepler-419 like evolution, before exiting the anti-aligned mode. A third category included in the pink circles is for example the 7:1 and 8:1 MMR for $M_d$ = 200 M$_{\rm J}$, where the system does evolve into K419-like state, but remains dynamically very noisy, even after the disk's dissipation. In this case, the disk contribution to the Hamiltonian is clearly of the same magnitude as the MMR.

The analytical Hamiltonian of PWA19 did not include any resonant terms, and these results indicates that even very high order MMRs can play an important role in exoplanetary dynamics. 

\begin{figure*}
\begin{centering}
	\includegraphics[scale=0.17]{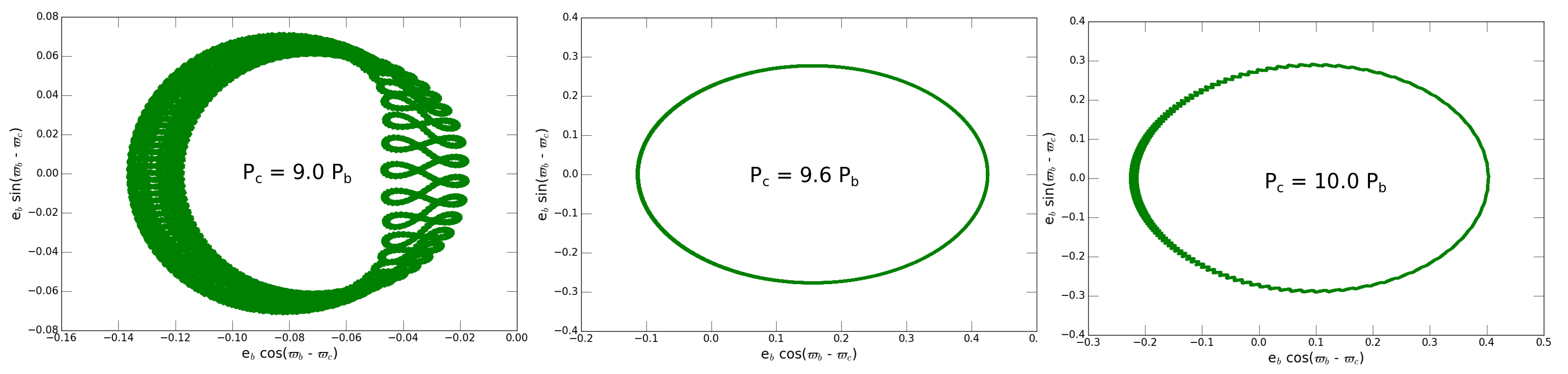}
   \caption{The evolution the system in the pseudo-coordinate-momentum phase space, over $8\sim\times10^3$ yr, starting at t=$2\times10^6$ yr for the 20 M$_J$ disk case (so the disk's potential is practically null). Note the different x and y scales used. }
    \label{fig:reso}
    \end{centering}
\end{figure*}

To check whether these variations are indeed caused by high order MMRs, we plot in Fig. \ref{fig:reso} the system short term evolution in the pseudo coordinate-momentum pair (e$_b$ cos[$\varpi_b$ - $\varpi_c$], e$_b$ sin[$\varpi_b$ - $\varpi_c$]) phase space for three cases: P$_c$ = 9.0, 9.6, and 10.0 P$_b$. We notice the appearance of new high frequency modes for the 9:1 and 10:1 MMR, and their complete absence outside of these resonances when P$_c$=9.6 P$_b$. The amplitude of these modes moreover seems to decrease with the increasing resonance order, implying furthermore that they are indeed MMRs related. Representative cases from this section were verified using the very high accuracy non-symplectic integrator \texttt{IAS15}.

In conclusion, for high enough disk masses, while there are large parts of P$_c$-M$_d$ parameter space where the system can evolve into Kepler-419, other end-results are also possible.
Two planets should not get trapped in certain higher order MMRs, or be close enough to undergo close encounters. Kepler-419 like systems might hence be more common than currently thought, and future observations can confirm or rule this out. The detailed eccentricity and apses evolution of all of these cases in shown in the appendix in Figures \ref{fig:app3} and \ref{fig:app4}.

\section{Discussions \& conclusions}
In this paper we studied the origins of the highly eccentric, apsidally anti-aligned, two giant planets Kepler-419 system using N-body simulations where we introduced a dissipating protoplanetary disk's potential as an extra force. We first show that the analytical results of PWA19 are recovered accurately, indicating that secular theory is adequate to describe the system's evolution with no significant effects from the ignored parts of the Hamiltonian. By exploring a large range of disk masses, we show that the minimum disk mass to retrieve Kepler-419 is 75 Jupiter masses, increasing to 95 Jupiter masses if we are to include the planets themselves. These values are consistent with values found in the infrared with Herschel, and lie towards the upper range of masses measured with ALMA, but CO based observations are probably underestimating the gas mass significantly. Furthermore, Kepler-419 is recovered even for dissipation timescales as low as $10^4$ yr, consistent with photoevaporation models. We then used a simple 1D $\alpha-$disk model to study the stability of a protoplanetary disk this massive, and find the Toomre $Q$ parameter to be significantly larger than unity for all reasonable radii. This imply that the Kepler-419 planets, massive as they are, probably formed via core accretion.

Finally we ran simulations while varying the period of planet $c$, and find that higher order MMRs such as the 8:1, 9:1, and 10:1 can either completely stabilize the system against the disk's precessional effects, or sometimes render its evolution dynamically noisy. 

{ Our results indicate that K419-like systems with highly eccentric apsidally anti-aligned planets might not be uncommon. Considering the set of all confirmed planets in the \texttt{exoplanets.eu} catalogue (as of August 2020), $\sim$ 7\% of all of the planet hosting stars have a mass $\geq$ 1.4 M$_\odot$. Since disk and stellar masses are strongly correlated, this can be considered as a very crude estimate for the prevalence of K419-capable disks in the population that surrounded these planet hosting stars. Moreover, in the 150 transition disks survey of \cite{vander}, a similar $\sim$ 6.6\% of the observed population was massive enough to account for K419. These numbers however are only reflective of the disks masses, and is certainly just a theoretical upper limit on the prevalence of K419-like systems, as only a fraction of these disks will stochastically follow this formation scenario. We emphasize that the stellar mass fraction is of the \textit{currently observed} planet-hosting stars, and not of the overall actual stellar (or planetary) population. These numbers, in addition to the results of section \ref{csm} showing that K419 could have formed from a wide range of period ratios, indicate that other apsidally anti-aligned systems could be hiding in the current confirmed planets catalogue. By searching for systems with two eccentric Jovian planets with P$_{out}$/P$_{in}$ $\geq$ 9 orbiting massive stars, we identify Kepler-432 and Corot-20 as prime candidates, followed by possibly Kepler-539.  }

Multiple simplifications were used in this work. We mainly did not take into account the feedback of the planets onto the disk. We assumed axi-symmetrical disks, and did not account for angular momentum exchange between the planets and disk \citep{lai}. While the disk's viscosity can damp an embedded planet's eccentricity \citep{bitsch1,bitsch2}, its Lindblad torques can excite it \citep{goldreich,duffel,teyssandier} (possibly explaining the intial AMD in the system). Recently \cite{lai} used an semi-analytic model to investigate secular dynamics of two planets systems in non axi-symmetrical discs, accounting for the planets eccentricity damping. They showed that, while for $\alpha=10^{-2}$ the dynamics we are considering here are suppressed, moderate eccentricity growth is possible for $\alpha=10^{-3}$. They did not explore the case of a quasi laminar ``dead zone'' disk with $\alpha=10^{-4}$ \citep{gammie}, a possibility consistent with recent ALMA observations \citep{turbobs1,turbobs2}. Moreover, their model and also ours did not account for planet eccentricity excitation due to the disk's torques, which likely operate for such massive planets ($m_c\simeq 7M_J$) through the 3:1 Lindblad resonance \citep{2001A&A...366..263P,bitsch1}. These aspects merit further investigation. 

Overall, our paper shows how unique exoplanetary systems architectures can be used to trace back the properties of the disk in which the planets formed. This is parallel to some exoplanets observed in mean motion resonances (GJ 876), for which disk-driven migration  captures depend on disk properties (scale height and levels of turbulence \citep{lee2002,rein2012,batygin}. Unlike the case of MMR captures however, the "capture" into the high-eccentricity secular equilibrium of Kepler-419 depends on the integrated evolution of the disk, placing a set of complementary constraints. {Future models taking into account the 3D architecture of the system \citep{petro2020} will also provide additional constrains. }

\section*{Acknowledgements}
We thank J. R. Touma at the American University of Beirut for interesting discussions that helped guide this project. We thank an anonymous referee for their insightful comments that helped improving this manuscript.  M.A.-D. is supported through a Trottier postdoctoral fellowship. The computations were performed on the Sunnyvale cluster at the Canadian Institute for Theoretical Astrophysics (CITA).

\section*{Data availability}
The data underlying this article (Rebound simulations binary archives) will be shared on reasonable request to the corresponding author.








\newpage

\begin{appendices}

\begin{figure*}
\begin{centering}
	\includegraphics[scale=0.30]{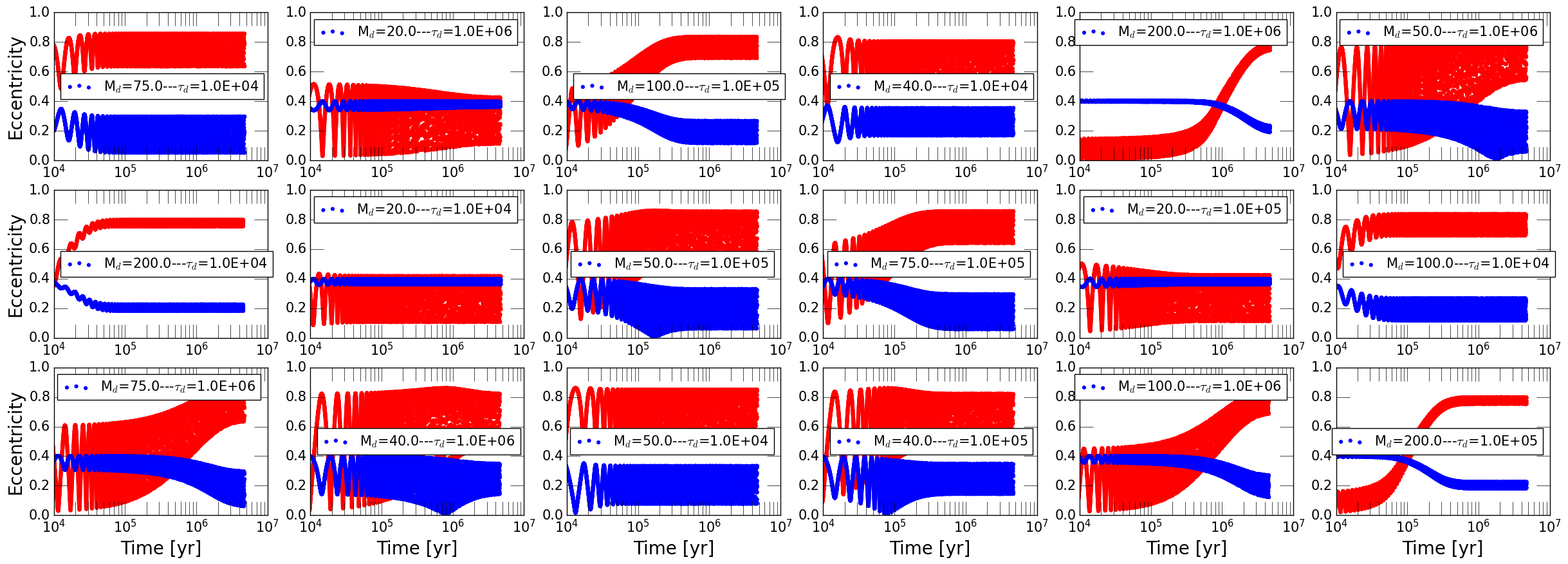}
   \caption{Appendix: Kepler-419 (P$_c$=9.6 P$_b$) planets eccentricity evolution for a wide range of disk mass and dispersal timescales. Red is for planet b, and blue is for planet c.}
    \label{fig:app1}
    \end{centering}
\end{figure*}

\begin{figure*}
\begin{centering}
	\includegraphics[scale=0.30]{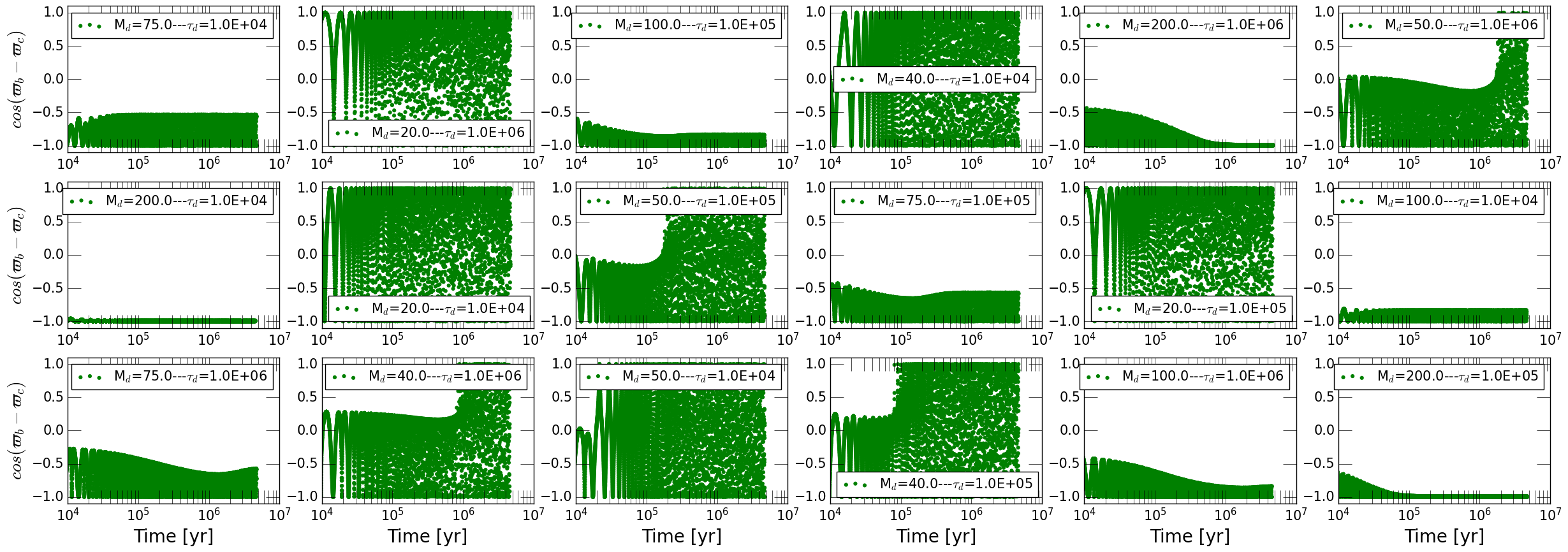}
   \caption{Appendix: Kepler-419 (P$_c$=9.6 P$_b$) planets apses evolution for a wide range of disk mass and dispersal timescales.}
    \label{fig:app2}
    \end{centering}
\end{figure*}

\begin{figure*}
\begin{centering}
	\includegraphics[scale=0.20]{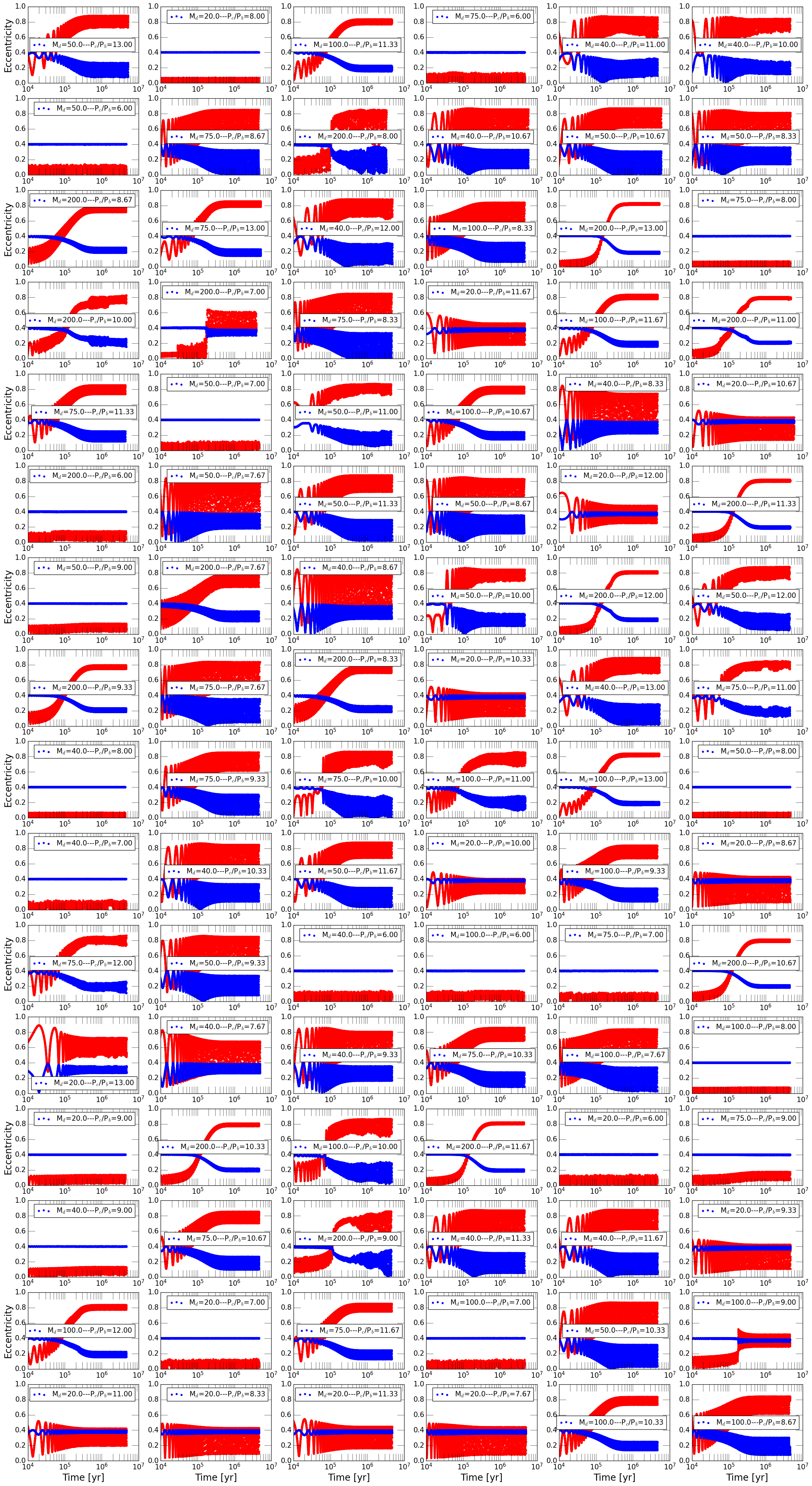}
   \caption{Appendix: Same as Fig. \ref{fig:app1}, but with different semi major axis values for planet $c$. Red is for planet b, and blue is for planet c. }
    \label{fig:app3}
    \end{centering}
\end{figure*}

\begin{figure*}
\begin{centering}
	\includegraphics[scale=0.20]{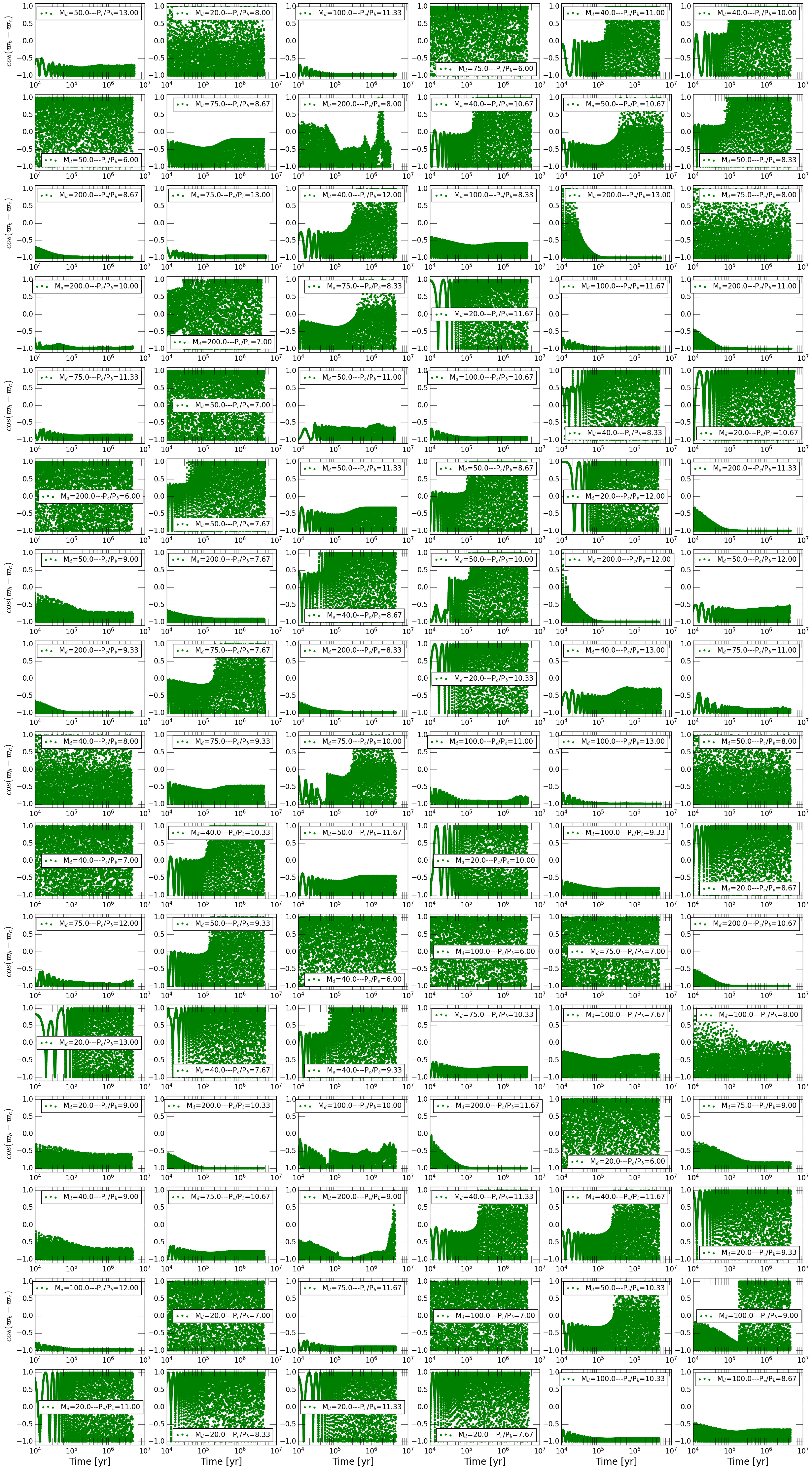}
   \caption{Appendix: Same as Fig. \ref{fig:app2}, but with different semi major axis values for planet $c$. }
    \label{fig:app4}
    \end{centering}
\end{figure*}

\end{appendices}


\bsp	
\label{lastpage}
\end{document}